\documentclass[12pt,preprint]{aastex}

\begin{document}
\title{MULTI-WAVELENGTH STUDY OF THE Be/X-RAY BINARY MXB~0656$-$072}
\author{Jingzhi Yan\altaffilmark{1,2,3}, Juan Antonio Zurita Heras\altaffilmark{2}, Sylvain Chaty\altaffilmark{2}, Hui Li\altaffilmark{1,3}, and Qingzhong Liu\altaffilmark{1,3} }
\altaffiltext{1}{Purple Mountain Observatory, Chinese Academy of Sciences, Nanjing 210008 , China; {\sf
jzyan@pmo.ac.cn,hli@pmo.ac.cn, qzliu@pmo.ac.cn}} \altaffiltext{2}{Laboratoire AIM, CEA/DSM-CNRS-Universit\'e Paris Diderot, IRFU/Service d'Astrophysique, FR-91191 Gif-sur-Yvette, France; {\sf juan-antonio.zurita-heras@cea.fr, sylvain.chaty@cea.fr}}\altaffiltext{3}{Key Laboratory of Dark Matter and Space Astronomy, Chinese Academy of Sciences, Nanjing 210008, China}

\begin{abstract}
We present and analyze the optical photometric and spectroscopic data of the Be/X-ray binary MXB~0656$-$072 from 2006 to 2009. A 101.2-day orbital period is found, for the first time, from the present public X-ray data (\emph{Swift}/BAT and \emph{RXTE}/ASM). The anti-correlation between the H$\alpha$ emission and the $UBV$ brightness of MXB~0656$-$072 during our 2007 observations indicates that a mass ejection event took place in the system. After the mass ejection, a low-density region might develop around the Oe star. With the outward motion of the circumstellar disk, the outer part of the disk interacted with the neutron star around its periastron passage and a series of the X-ray outbursts were triggered between MJD~54350 and MJD~54850. The PCA--HEXTE spectra during the 2007-2008 X-ray outbursts could be well fitted by a cut-off power law with low energy absorption, together with an iron line around 6.4~keV, and a broad cyclotron resonance feature around 30~keV. The same variability of the soft and hard X-ray colors in 2.3-21~keV indicated that there were no overall changes in the spectral shape during the X-ray outbursts, which might be only connected with the changes of the mass-accretion rate onto the neutron star.

\end{abstract}
\maketitle
\section{Introduction}
Be/X-ray binaries represent the largest subclass of high mass X-ray binaries, in which a neutron star orbits around a massive and early type star in a wide and eccentric orbit. Currently, no Be/black-hole system has been found \citep{Zhang04}. There are two different kinds of disks in Be/X-ray binaries, a circumstellar disk around the Be star and an accretion disk around the neutron star. The formation of a Be disk may be connected with the rapid rotation of the Be star \citep{Porter03}. Be/X-ray binaries show two different X-ray outburst activities: Type I X-ray outbursts, which are usually associated with the periastron passage of the neutron star through the environment of the circumstellar disk, and Type II X-ray outbursts, which could occur at any orbital phase and last for a large fraction of the orbital period (see \citet{Reig11} and the references therein).

The transient X-ray source MXB~0656$-$072 was discovered on 1975 September 20 with SAS-3 \citep{Clark75}. The following high-sensitivity-mode observations showed that the 3--6~keV flux of the source was 0.05 and 0.07 times that of the Crab Nebula on 1976 March 19 and March 27, respectively \citep{Kaluzienski76}. According to its long-term variability in X-rays, MXB~0656$-$072 was classified as a low mass X-ray binary by \citet{Liu01}.

\emph{RXTE}/ASM detected a new X-ray outburst of MXB~0656$-$072 in 2003 October \citep{Remillard03} and reached a peak X-ray luminosity of 200~mCrab. \emph{RXTE}/PCA observations on 2003 October 19--20 indicated that the source was a pulsar with a spin period of 160.7~s \citep{Morgan03}. Using ROSAT PSPC observations, \citet{Pakull03} identified the optical counterpart and revealed an O9.7\,Ve spectral type. Thus, MXB~0656$-$072 was classified as a Be/X-ray binary in a recent high mass X-ray binary catalog \citep{Liu06}.
\citet{McBride06} analyzed the \emph{RXTE} observations during the 2003 outburst and found a cyclotron resonance  scattering feature
at 32.8~keV and an average pulse period of 160.4$\pm$0.4~s with a spin-up of 0.45~s across the outburst. They also estimated a distance of 3.9$\pm$0.1~kpc to the source from the observed $V$ magnitude ($V$ = 12.05--12.38) and $B-V$ color ($B-V$=1.02--0.86) \citep{Pakull03}.

A series of X-ray outbursts that occurred between 2007 November 10 and 2008 November 10 was reported by \emph{International Gamma-Ray Astrophysics Laboratory} \citep{Kreykenbohm07}, \emph{RXTE}/ASM \citep{Pottschmidt07} and \emph{Swift}/BAT \citep{Kennea07}. \citet{Yan07a} reported their spectroscopic observations on MXB~0656$-$072 and found a strong H$\alpha$ emission line before its 2007 X-ray outburst.

In this paper, we present our optical photometric and spectroscopic observations on MXB~0656$-$072 from 2005 to 2009 and simultaneous X-ray observations in Section~\ref{SecObs}. The optical and X-ray variability during different X-ray states is analyzed in Section~\ref{SecAna}. In Section~\ref{Disscussion}, we discuss the X-ray periodic outburst nature of the system. The results are summarized in Section~\ref{SecSum}.

\section{Observations}\label{SecObs}

The multi-wavelength observations, including the intermediate-resolution optical spectroscopy, $UBV$ photometry, and X-rays, are described in the following subsections.

\subsection{Optical Spectroscopy}

We obtained several spectra of MXB~0656$-$072 with the 2.16~m telescope at Xinglong Station of National Astronomical Observatories, China (NAOC), from 2005 to 2009. The optical spectroscopy with an intermediate resolution of 1.22~{\AA}\,pixel$^{-1}$ was performed with a CCD grating spectrograph at the Cassegrain focus of the telescope. We took blue and red spectra covering the wavelength ranges 4300--5500 and 5500--6700\,{\AA}, respectively, at different times. All spectra were reduced with the IRAF\footnote{IRAF is distributed by NOAO, which is operated by the Association of Universities for Research in Astronomy, Inc., under cooperation with the National Science Foundation.} package. The data were bias-subtracted and flat-field-corrected, and the cosmic rays were removed. Helium-argon spectra were taken in order to obtain the pixel--wavelength relation. To improve this relation, we also used the diffuse interstellar bands at 6614 and 6379~{\AA} observed in the spectra.

Some selected and normalized red spectra from 2005 to 2009 and a zoom of the H$\alpha$ region are shown in Figures~\ref{figure:spectra}~(a) and (b), respectively. The corresponding observational dates are marked in the left part of each spectrum. The equivalent width (EW) of the H$\alpha$ line has been measured by selecting a continuum point on each side of the line and integrating the flux relative to the straight line between the two points using the procedures available in IRAF. The measurements were repeated five times for each spectrum and the error was estimated from the distribution of obtained values. The EW(H$\alpha$) typical error is within 3\%. This error arises due to the subjective selection of the continuum. The EW(H$\alpha$) results are listed in Table~\ref{table_halpha} together with the observational date, the MJD, and the exposure.

\begin{center}
\begin{figure}
\centering
\includegraphics[width=12cm]{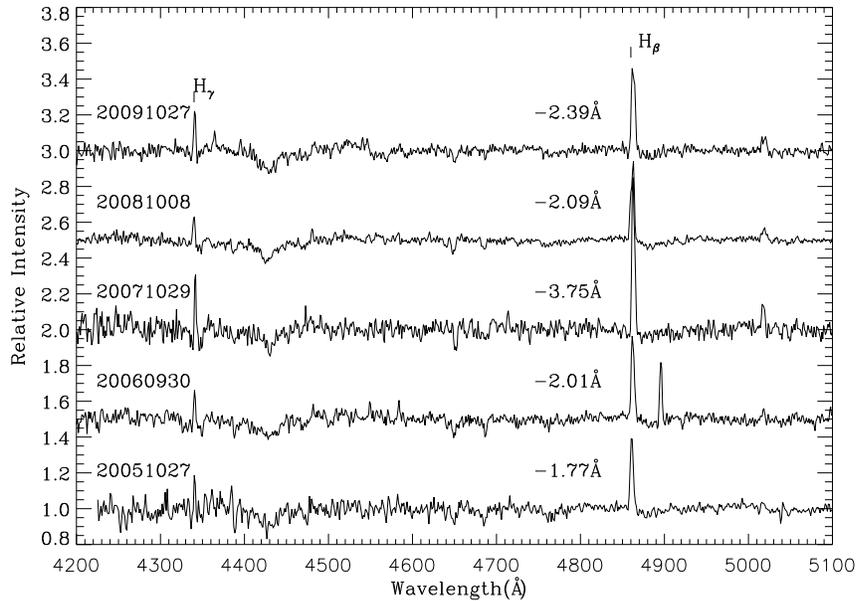}%
\caption{Blue spectra of MXB~0656$-$072 from 2005 to 2009. All spectra have the continuum level normalized and offset vertically to allow direct comparison. Every observational date and EW(H$\beta$) are reported for each spectrum. The spike near the H$\beta$ line on 2006 September 30 is caused by a cosmic ray.}
\label{figure:blue}
\end{figure}
\end{center}

The blue spectra were as well taken during our five years of observations (see Fig.~\ref{figure:blue}). HeI~$\lambda$5016, H$\beta$, and H$\gamma$ are clearly seen. Both observational date and EW(H$\beta$) are marked at the left and right part of each spectrum, respectively.

\subsection{$UBV$ Photometry}

Since 2007, we performed systematic photometric observations on a sample of X-ray binaries with both the 100~cm Education and Science Telescope (EST) and the 80~cm Tsinghua-NAOC Telescope (TNT) at Xinglong Station of NAOC. The EST, manufactured by EOS Technologies, is an altazimuth-mounted reflector with Nasmyth foci at a focal ratio of $f/8$. TNT is an equatorial-mounted Cassegrain system with a focal of $f/10$, made by AstroOptik, funded by Tsinghua University in 2002 and jointly operated with NAOC. Both telescopes are equipped with the same type of Princeton Instrument 1340$\times$1300 thin back-illuminated CCD. The CCD cameras use standard Johnson-Cousins $UBVRI$ filters made by Custom Scientific. Table~\ref{table:logphotometry} reports the log of our photometric observations performed on MXB~0656$-$072 using $UBV$ filters, including the observational date, the number of frames and exposure time per filter, and the seeing.

\begin{center}
\begin{figure}
\centering
\includegraphics[width=12cm]{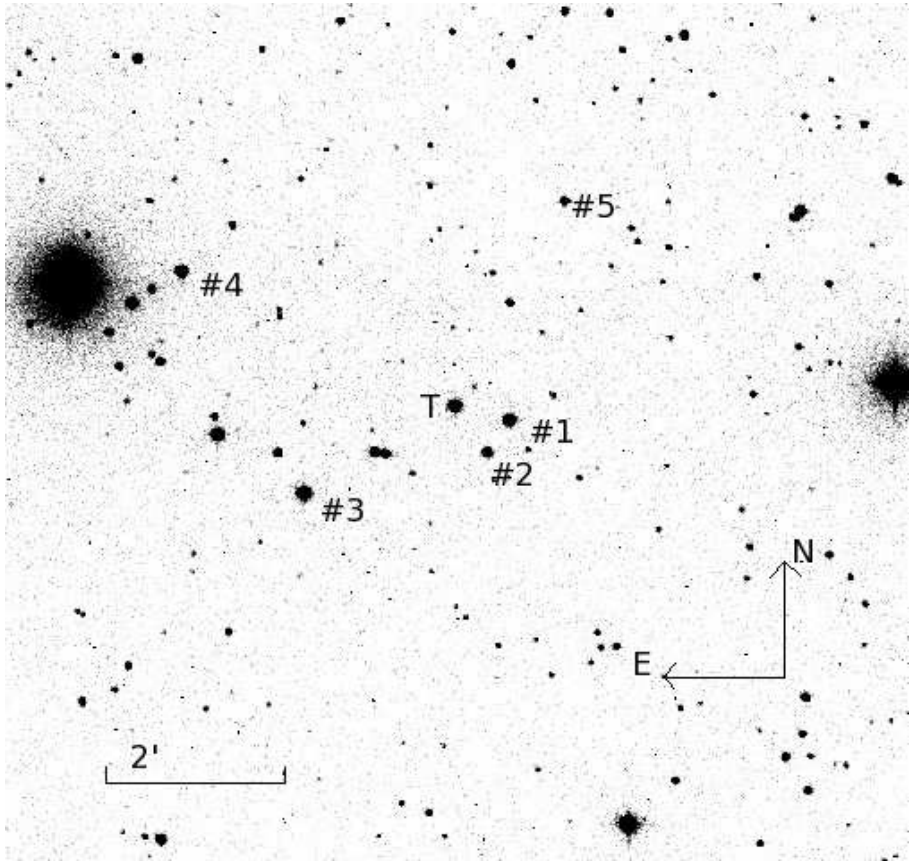}%
\caption{B-band TNT image of MXB~0656$-$072 taken on 2008 October 08. MXB~0656$-$072 is marked with 'T' and five reference stars are marked with \#1 to \#5.}  \label{figure:field}
\end{figure}
\end{center}

The photometric data reduction was performed using standard routines and aperture photometry packages in IRAF, including bias subtraction and flat-field correction. In order to study the variation of the $UBV$ brightness, we selected five reference stars in the field of view (see Fig.~\ref{figure:field}) to derive the differential magnitude of MXB~0656$-$072. The results indicate that the reference star \#3 is more stable than the other four reference stars. Thus, we selected star \#3 as the main reference star and star \#4 as check star. The $UBV$ differential magnitudes and errors of MXB~0656$-$072 in our 2007, 2008, and 2009 observations are listed in Table~\ref{table:diffmag}. The errors are calculated as the square root of the sum of the squares of the measured errors of the target and the main reference star.
The typical error for $UBV$ differential magnitudes is within 1\%, except for the data on 2009 October 27, when the larger errors may have been due to the bad seeing.

\subsection{X-Ray Observations}

The All Sky Monitor (ASM, 1.5--12~keV) on board \emph{RXTE} \citep{Levine96} and the Burst Alert Telescope (BAT, 15--50~keV) on board \emph{Swift} \citep{Barthelmy00} monitored the X-ray activity of MXB~0656$-$072 since 1998 July and 2005 February, respectively. The X-ray light curves of both \emph{RXTE}/ASM and \emph{Swift}/BAT cover a series of X-ray outbursts since 2007 November (see Figure~\ref{figure:asmbatlc}). The positions of the vertical bars at the bottom of the ASM light curve correspond to the beginning time of the \emph{RXTE}-pointed observations.

\emph{RXTE} public data during the outbursts of MXB~0656$-$072 between 2007 November and 2008 November were retrieved from the data archive of the High Energy Astrophysics Science Archive Research Center (HEASARC). The \emph{RXTE}-pointed observations with observational ID~93032 and 93423 have a total effective exposure time of $\sim407$~ks. The \emph{RXTE}/PCA standard2 data with a time-resolution of 16~s were used to extract the light curve and the spectrum of the source in the energy range of 3--22~keV. Only Proportional Counter Unit 2 (PCU2) was used for the analysis, because it was always operational during all the pointed observations of MXB~0656$-$072. We extract Proportional Counter Array (PCA) light curves and spectra when the source has an offset angle lower than 0\fdg02 and when the earth limb is situated more than 10\degr\ away of the source direction. The latest version of PCA background models, which were available at HEASARC Web site, was used to generate the background light curves and spectra.

The standard mode data of HEXTE cluster B are also used in the analysis. HEASoft script {\tt hxtback} is used to separate the background and source files from the raw FITS files. HEXTE spectra are extracted using {\tt saextrct}, similar to the PCA standard2 data reduction, from the detectors of SpecDet0, SpecDet1, and SpecDet3 in cluster B. Dead time of each spectra is corrected using {\tt hxtdead}. The background and source spectra are then combined using {\tt sumpha}. The response files are generated and added into the spectra files using the script {\tt hxtrsp}.

\section{Analysis}\label{SecAna}

\subsection{Periodic Variation in X-Ray Light Curves}\label{subperiod}

\begin{center}
\begin{figure}
\centering
\includegraphics[width=12cm]{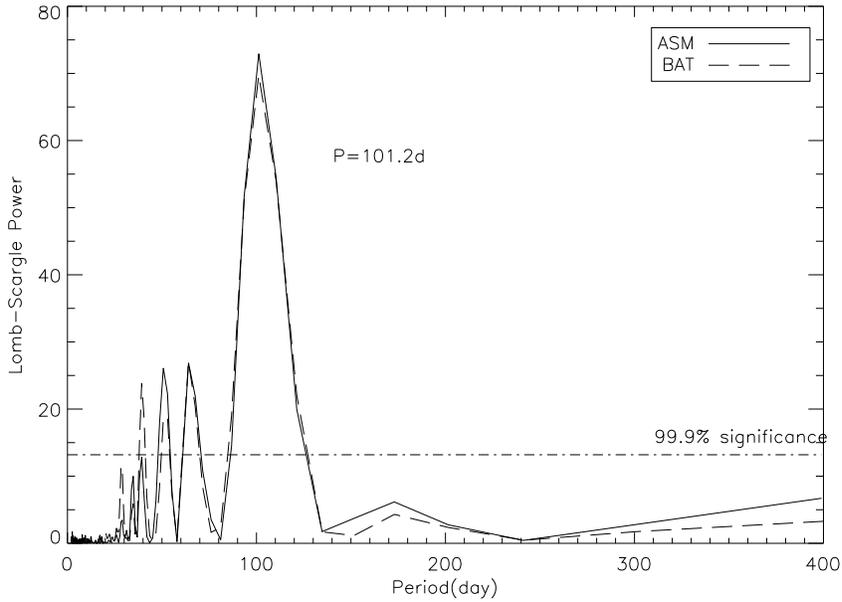}%
\caption{Lomb-Scargle periodogram of ASM (solid) and BAT (dashed) light curves for the source of MXB~0656$-$072. The horizontal dot-dashed line is the 99.9\% confidence level.}  \label{figure:lomb}
\end{figure}
\end{center}

\begin{center}
\begin{figure}
\centering
\includegraphics[width=12cm]{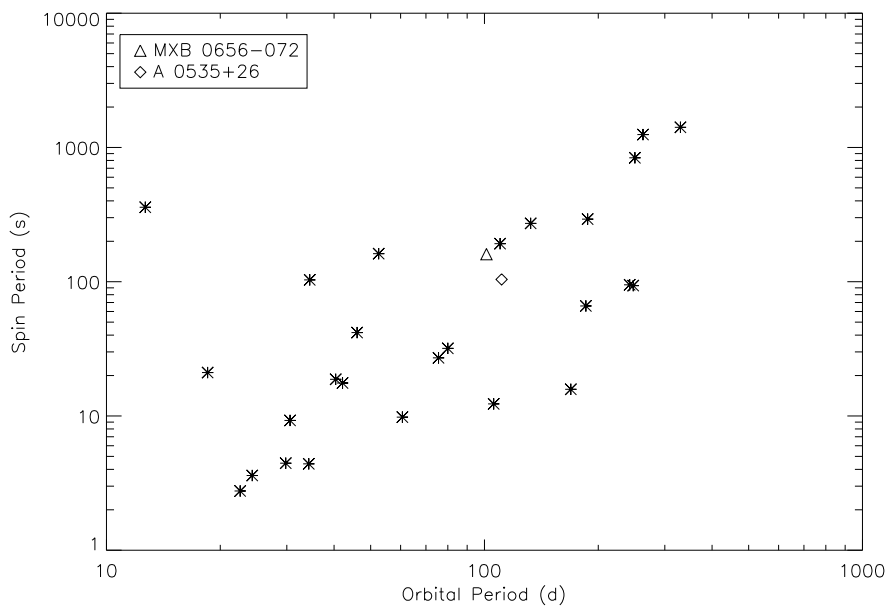}%
\caption{Position of MXB~0656$-$072 (triangle) in the Corbet diagram. The asterisks are Be/X-ray binaries in catalogs of \citet{Liu06} and \citet{Raguzova05}. The diamond corresponds to the position of Be/X-ray binary A0535+262.}  \label{figure:corbet}
\end{figure}
\end{center}

Four continuous X-ray outbursts separated by $\sim100$ days were detected by \emph{RXTE}/ASM and \emph{Swift}/BAT telescopes (see Figure~\ref{figure:asmbatlc}). These series of outbursts are similar to the periodic Type I X-ray outbursts at periastron passage in Be/X-ray binaries. We selected both X-ray light curves between MJD 54350--54850 to search for periodic variability using the Lomb-Scargle (LS) periodogram method \citep{Scargle82}. A clear peak at 101.2 days is visible for both ASM (solid line) and BAT (dashed line) LS periodograms (see Figure~\ref{figure:lomb}). Combining this value with the spin period of 160.4~s in the $P_{\mathrm{orb}}-P_{\mathrm{spin}}$ diagram \citep{Corbet86}, the source is indeed located in the region of Be/X-ray binaries (see Figure~\ref{figure:corbet}). The parameters of the spin and orbital periods of Be/X-ray binaries in Figure~\ref{figure:corbet} are from \citet{Liu06} and \citet{Raguzova05}. The position of MXB~0656$-$072 in the Corbet diagram  is close to that of the Be/X-ray binary A0535$+$262 (diamond in Figure~\ref{figure:corbet}), which has an orbital period of $P_{orb}$ $\sim$ 111.1 days and a spin period of $P_{spin}$ $\sim$ 103s \citep{Finger96}. The 101.2-day period should be the orbital period of MXB~0656$-$072 and it might produce observational phenomena similar to those of the Be/X-ray binary A0535$+$262.

\begin{center}
\begin{figure}
\centering
\includegraphics[width=12cm]{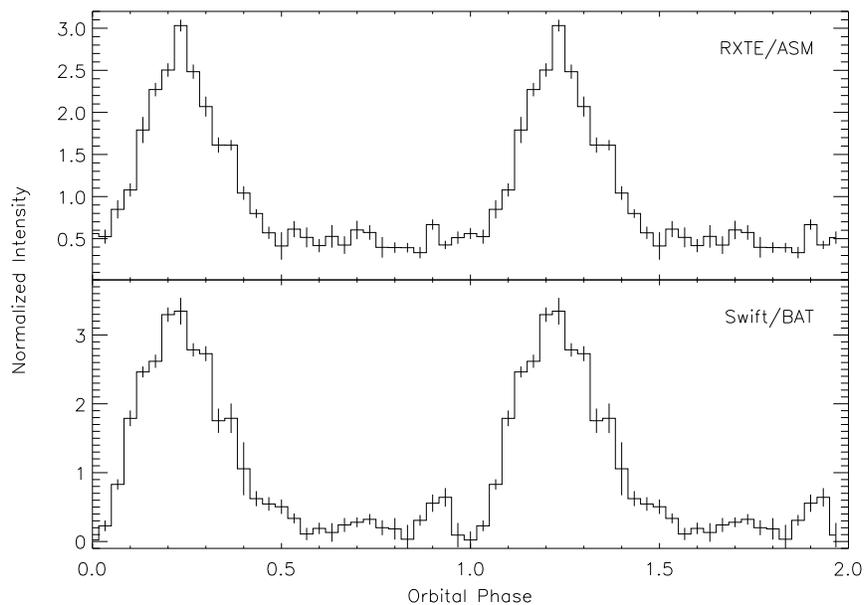}%
\caption{ASM and BAT folded light curves at the period of $P_{orb}$=101.2 days. The zero phase epoch (MJD~54408) corresponds to the beginning of the first (2007) outburst.}  \label{figure:foldlc}
\end{figure}
\end{center}

Both ASM and BAT light curves in Figure~\ref{figure:foldlc} are folded with a period of $P=101.2$~d and a zero phase epoch of MJD~54408 that corresponds to the beginning of each outburst.

\subsection{Optical Variability}

\begin{center}
\begin{figure}
\centering
\includegraphics[width=8cm]{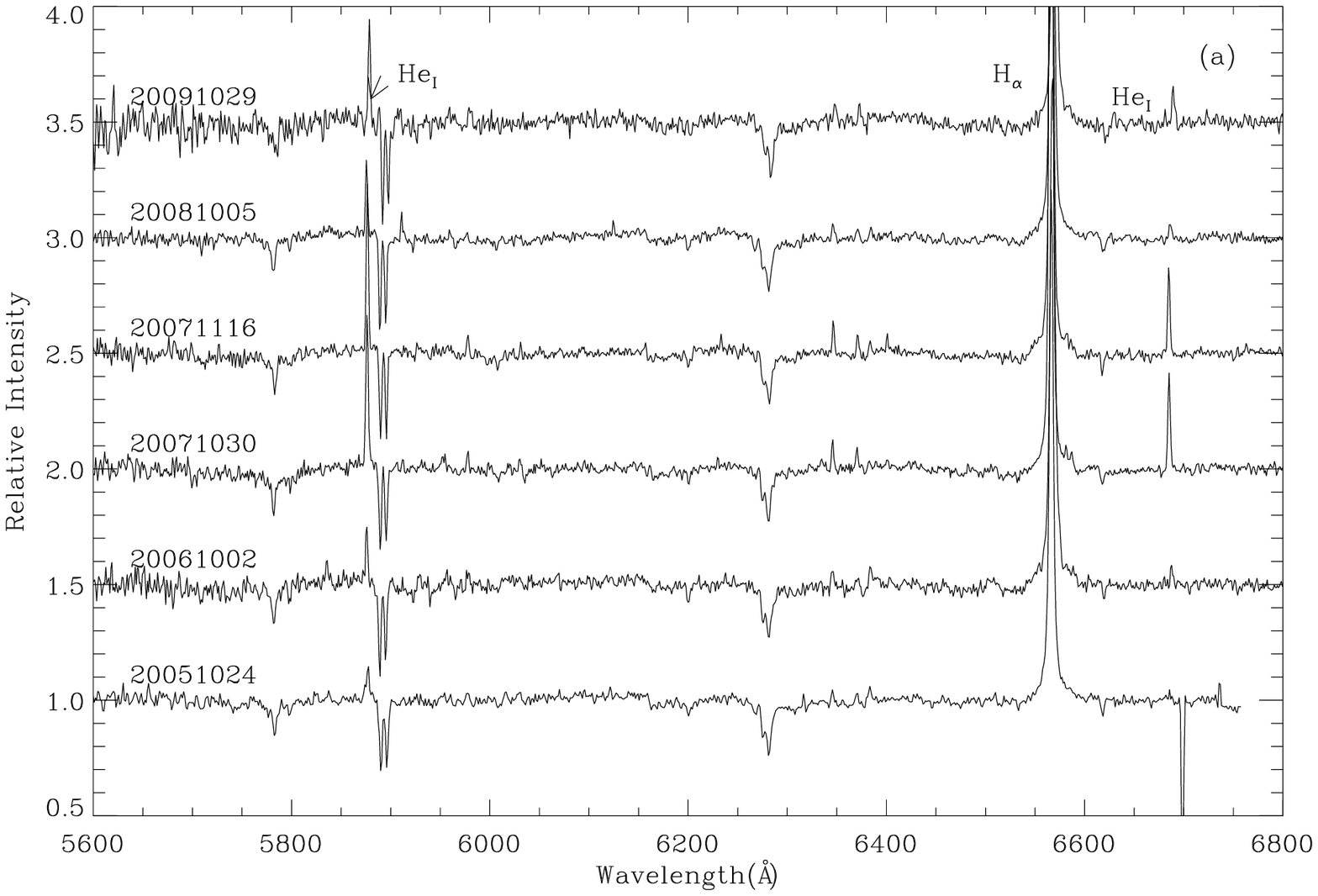}%
\includegraphics[width=8cm]{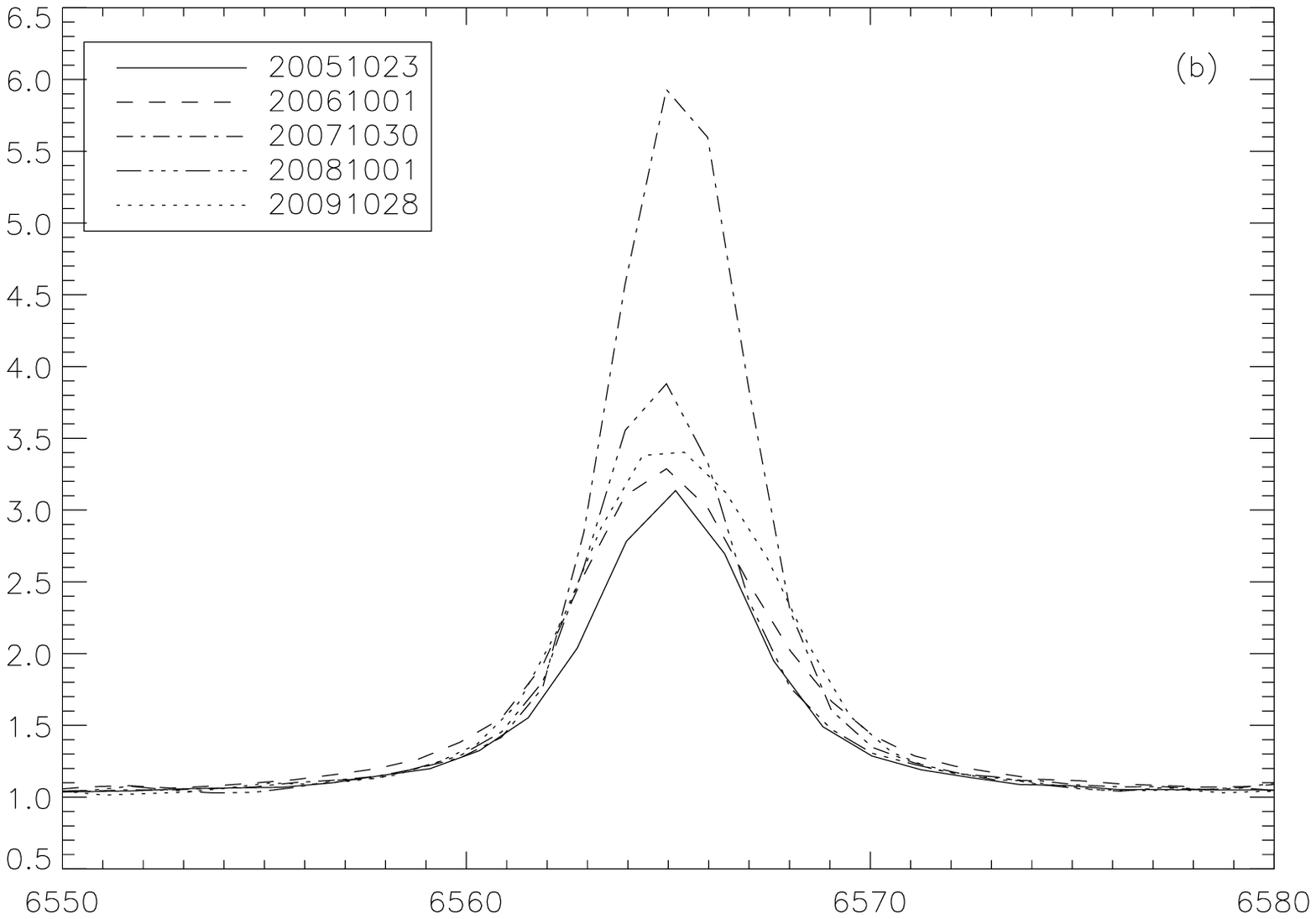}%
\caption{(a) Selected spectra from our 2005 to 2009 observations. All spectra have the continuum level normalized and offset vertically to allow direct comparison. The observational date is marked on the left part of each spectrum in the YYYYMMDD format. The dip on the right part of spectrum on 2005 October 24 was caused by bad pixels on the CCD.  (b) The zoomed panel of the left figure in the H$\alpha$ region with all spectra at the same continuum level.}  \label{figure:spectra}
\end{figure}
\end{center}

\begin{center}
\begin{figure}
\centering
\includegraphics[height=12cm]{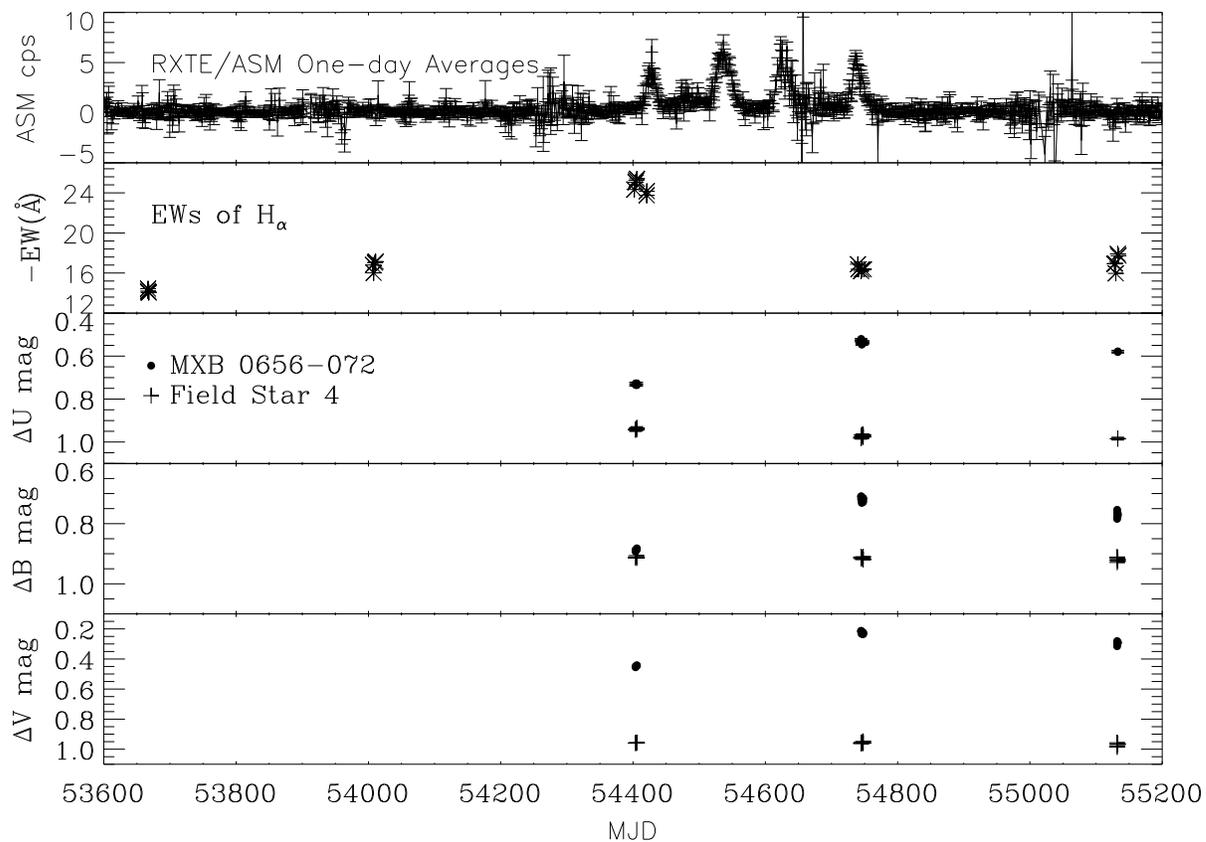}%
\caption{Evolution of EW of H$\alpha$ and $UBV$ differential magnitudes of MXB~0656$-$072, together with the \emph{RXTE}/ASM X-ray light curve in 1.5-12.0~keV. The $UBV$ differential magnitudes of check star 4 are also plotted in the figure. The errors for the EW of H$\alpha$ and $UBV$ differential magnitudes are smaller than the sizes of the corresponding symbols in the figure.}  \label{figure:uvb}
\end{figure}
\end{center}

Normalized optical spectra of MXB~0656$-$072 in red and blue regions are shown in Figure~\ref{figure:spectra} and Figure~\ref{figure:blue}, respectively. H$\alpha$, HeI~$\lambda\,$5875,6678, H$\beta$, and H$\gamma$ are clearly detected. The H$\alpha$ line shows a single-peaked and narrow profile during our observations. All emission lines became very strong just before the X-ray outburst in 2007 November. We plot the EWs(H$\alpha$), the $UBV$ differential magnitudes, and the 1.5--12~keV \emph{RXTE}/ASM light curve in Figure~\ref{figure:uvb}. Due to the larger error in the $U$-band magnitude, we do not plot the data of 2009 October 27 (MJD~55131) in Figure~\ref{figure:uvb}.

The results show that the strength of the H$\alpha$ line in our 2006 observations became stronger than that of 2005 and it had an extraordinary strength during our 2007 observations, which were taken just before the first X-ray outburst of 2007 November. Our optical spectroscopic observations on 2007 November 16 indicated that the flux of H$\alpha$ kept nearly constant during the X-ray outburst. In comparison with the H$\alpha$ emission lines in our 2005 and 2006 observations, they had almost the same width of the line wings (see Figure~\ref{figure:spectra}(b)), but the H$\alpha$ line during the 2007 observations became much stronger. The H$\alpha$ flux during our 2008 observations dropped to the same level as that in 2006. The optical spectroscopic observations in 2008 were carried out just during the last X-ray outburst. The strength of H$\alpha$ emission line during our 2009 observations changed little with respect to the strength in 2008. The H$\beta$, H$\gamma$, and He I lines also had the same variability: for instance, the He~I~$\lambda$\,6678\,{\AA} became very strong during our 2007 observations (with an average EW of $\sim$-1.34~{\AA}) and nearly lost its emission feature in 2005 and 2006 (see Figure~\ref{figure:spectra}(a)). In our 2008-2009 observations, the He~I~$\lambda$\,6678\,{\AA} line became faint again, but still stronger than in 2005 and 2006.

In 2007, our simultaneous optical photometry and spectroscopy on MXB~0656$-$072  showed an interesting behavior (see Figure~\ref{figure:uvb}): while the H$\alpha$ emission line strongly increased in 2007, the source brightness in $UBV$ decreased by 0.2 mag in 2007 compared to the 2008--2009 observations.

\subsection{\textbf{X-Ray Spectral Properties during the X-Ray Outbursts}}

\begin{center}
\begin{figure}
\centering
\includegraphics[height=12cm]{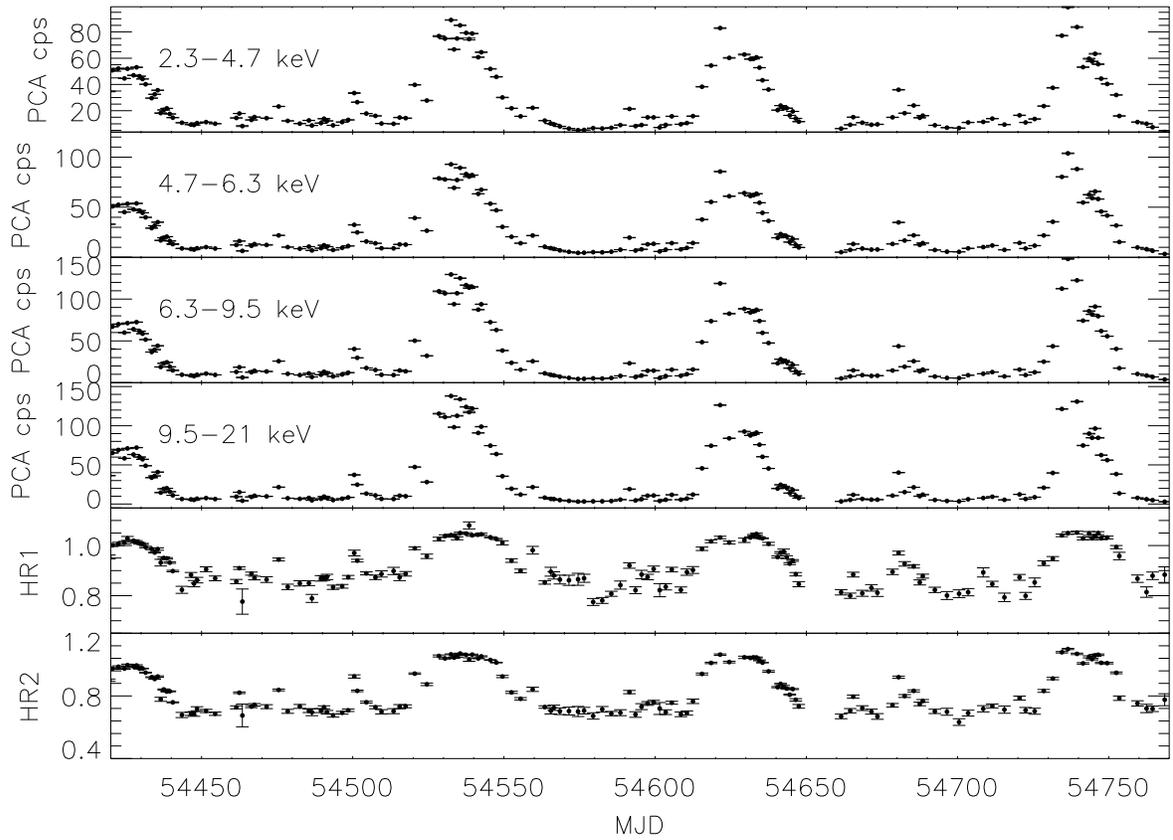}
\caption{PCA light curves in four different energy bands. The soft and hard colors, defined as HR1$\equiv$4.7--6.3~keV/2.3--4.7~keV and HR2$\equiv$9.5--21.0~keV/6.3--9.5~keV, respectively, are shown at the bottom. The size of some error bars is smaller than the size of the symbols.}
\label{figure:hr}
\end{figure}
\end{center}

We extract light curves from PCA standard2 data in four different energy bands: 2.3--4.7~keV, 4.7--6.3~keV, 6.3--9.5~keV, and 9.5--21~keV. The soft and hard colors are defined as the count rate ratios 4.7--6.3~keV/2.3--4.7~keV and 9.5--21.0~keV/6.3--9.5~keV, respectively. PCA light curves and the two colors are plotted in Figure~\ref{figure:hr}. The X-ray outbursts are also evident in PCA light curves. The soft and hard colors show the same variability during the X-ray outbursts.

We selected a typical observational ID (ObsID) 93032-30-01-03, which was taken around the time of the peak flux of the third outburst in Figure~\ref{figure:asmbatlc} (the vertical arrow at the bottom of the top panel), to explore the X-ray spectral properties during an X-ray outburst.
We modeled the PCA (3--22~keV) and HEXTE (18--80~keV) spectra using ISIS \citep[v1.6.1,][]{Houck2000}. Energies lower than 3~keV were ignored due to uncertainties in the background modeling and a systematic error of 0.5\% was added in quadrature to the PCA data in order to take the uncertainties in the response matrix into account. To improve the signal to noise in each channel, we grouped the PCA and the HEXTE spectra to a minimum signal to noise of 5 above a lower bound of 3~keV and 18~keV, respectively.

There exists no convincing theoretical model for the shape of the continuum X-ray spectrum in accreting X-ray pulsars. Their X-ray spectrum is generally modeled as a power law with high energy exponential cutoff.  We first tried to fit the PCA-HEXTE spectrum of MXB~0656$-$072 using a standard ISIS model {\tt cutoffpl}, a power law with a high-energy exponential cutoff,
 \begin{equation}
\label{euqation_cutoffpl}
I_{cont}(E)=\alpha E^{-\Gamma}e^{-E/E_{cut}},
\end{equation}
where $\alpha$ is the photons $keV^{-1}~cm^{-2}~s^{-1}$ at 1~keV, $\Gamma$ is the power law photon index, and $E_{cut}$ is the e-folding energy of exponential cutoff.

Another similar continuum model we tried is a power law with a high-energy rollover \citep[PLCUT;][]{White83}, which was applied to MXB~0656-072 during the 2003 X-ray outburst by \citet{McBride06}. The analytic form of the PLCUT model is
 \begin{equation}
\label{euqation_plcut}
I_{cont}(E)=\alpha\times E^{-\Gamma}\left\{ \begin{array}{ll}
1 & \textrm{$E \leq E_{cut}$}\\
e^{-(E-E_{cut})/E_{fold}} & \textrm{$E > E_{cut}$,}
\end{array} \right.
\end{equation}
where $E_{cut}$ and $E_{fold}$ are the cutoff and folding energies, respectively.

\begin{center}
\begin{figure}
\centering
\includegraphics[width=10cm]{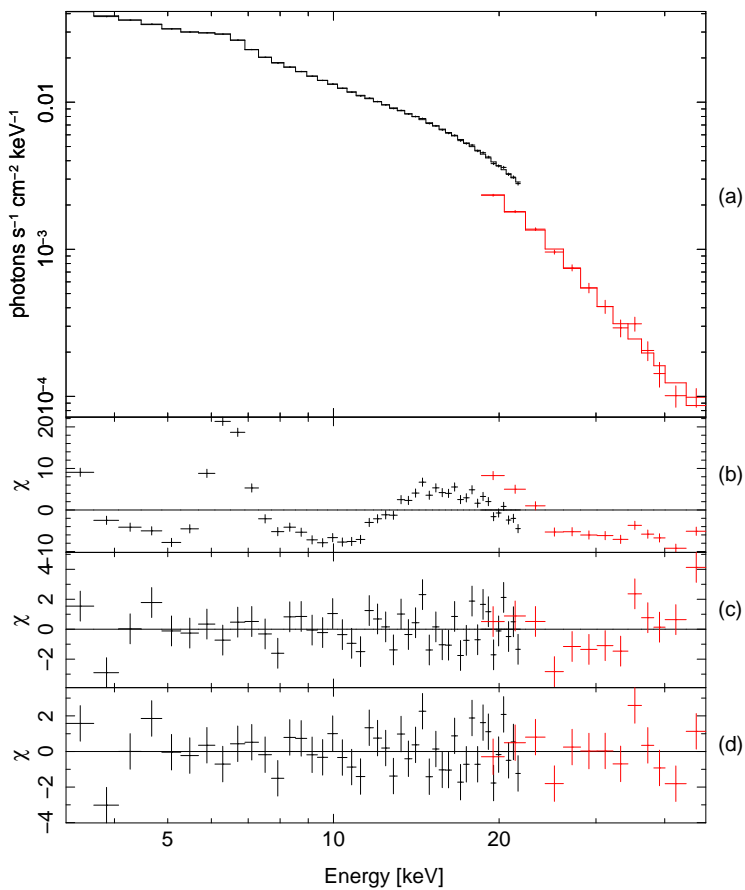}%
\caption{Best-fit PCA and HEXTE spectra for the ObsID 93032-30-01-03 modeled with a cutoff power law with low-energy absorption, a Gaussian line around 6.4~keV and a cyclotron resonance scattering feature around 30~keV. The panels (b), (c), and (d) are the residuals only for the {\tt cutoffpl} with the low energy absorption, the residuals without {\tt cyclabs}, and the best-fit residuals, respectively. The residuals of the fit are in terms of $\sigma$ values.} \label{figure:fit}
\end{figure}
\end{center}

\begin{center}
\begin{figure}
\centering
\includegraphics[width=10cm]{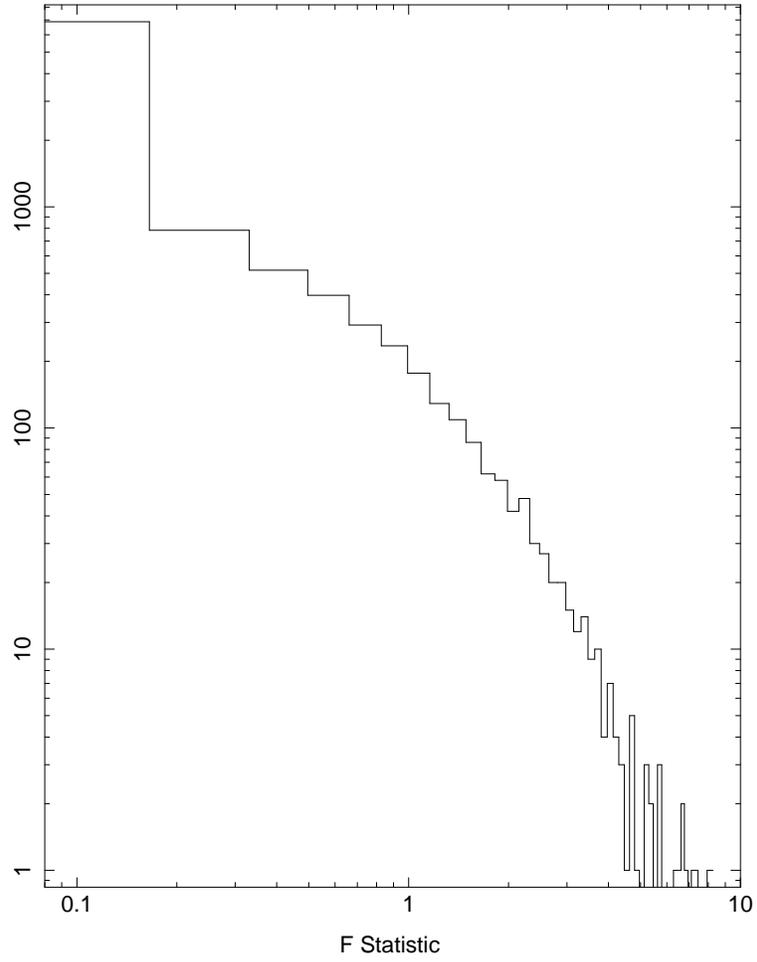}%

\caption{F-statistic distribution between the null model (see the text) and an alternative model with the addition of a gaussian line around 6.4~keV for 10000 simulated spectra.} \label{figure:monte}
\end{figure}
\end{center}

Each fit shows in the residuals a pseudo-P Cygni profile near 10~keV, which has been reported by \citet{McBride06} and found in other accreting X-ray pulsars \citep[Her X-1, GS 1843+00;][]{Coburn02}.  They concluded that the pattern of the residuals around 10~keV was due to the continuum model. In the PLCUT model, the energy cutoff should be smoothed by including a Gaussian absorption line with an energy dependent width at the continuum cutoff energy \citep{McBride06}. Here we choose to use {\tt cutoffpl} with a low-energy absorption, as it is the simplest model, to fit the X-ray continuum emission for the spectrum of MXB~0656-072. A normalization constant (1 for the PCA data and a free parameter for the HEXTE data) was added to the model to get a good agreement between these two instruments. A bad reduced Chi-squared $\chi^2_\nu$ = 21.5 for 53 degrees of freedom (dof) was obtained and the residuals are shown in the panel (b) of Figure~\ref{figure:fit}. Since the feature around 10~keV might not be real, we will not discuss its physical nature in this paper. In order to remove the systematic feature in the residuals at $\sim$10~keV, a Gaussian absorption line ({\tt gabs}) function,
 \begin{equation}
\label{euqation1}
GABS(E)=exp(-\frac{\tau}{\sqrt{2\pi}\sigma})\times exp(-\frac{(E-E_{l})^{2}}{2\sigma^{2}}),
\end{equation}
 was used in the model and improved the fit from a reduced Chi-square of 21.5 for 53 dof to 8.9 for 50 dof, where $\tau$ is the optical depth, $\sigma$ is the line width, and $E_l$ is the line energy. The other obvious feature in the panel (b) of Figure~\ref{figure:fit} should be an iron line at $\sim$ 6.4~keV, which was described by the Gaussian shaped model {\tt gaussian} in ISIS. Due to the lower energy resolution of PCA (less than 18\% at 6~keV), the width of the iron line in the model was fixed at 0~keV. After this feature was added in the model, a better reduced Chi-squared of 1.23 for 48 dof was obtained. Thus, the F-statistic is 156.7 and the F-test probability is about $9.1\times10^{-22}$, which indicates that it is reasonable to add a Gaussian component in the model. \citet{Protassov02} suggested that the F-test statistic cannot be used to test for the presence of a line and they proposed a Monte Carlo simulation approach to calculate the approximate statistical significance. We followed this method to calibrate the sample distribution of the F-statistic for the spectrum of MXB~0656-072. We simulated 10,000 spectra with ISIS from the null model (a cutoff power law with the low-energy absorption and a Gaussian absorption feature around 10~keV) at its best-fit parameters. Each faked data were grouped exactly as we did for the real data and fitted with the alternative model (which included the Gaussian line component). The F-statistic between the null and the alternative model was calculated and their distribution was shown in Figure~\ref{figure:monte}. The maximum value of the F-statistic from the 10,000 faked spectra is much less than the F-statistic value of 156.7 calculated from the real data. Therefore, we concluded that the ion line around 6.4~keV was detected at the greater than 99.99\% confidence level in the spectrum of MXB~0656-072.

We found a broad absorption feature around 30~keV in the residuals of the last fitting which showed in the panel (c) of Figure~\ref{figure:fit}. This feature should be the cyclotron resonance scattering feature around $\sim$30--40~keV reported by \citet{McBride06} and it was modeled with the {\tt cyclabs} function in ISIS,
 \begin{equation}
\label{euqation1}
CYCLABS(E)=exp(-D_{cycl}\frac{(W_{cycl}E/E_{cycl})^2}{(E-E_{cycl})^2+{W_{cycl}}^2}),
\end{equation}
where, $E_{cycl}$, $D_{cycl}$, and $W_{cycl}$ are the cyclotron energy, depth, and width, respectively \citep{Mihara90,Makishima90}. Here, the cyclotron width $W_{cycl}$ was fixed at 10~keV in the model. A good reduced Chi-square of 1.01 for 46 dof was obtained and the residuals were shown in the panel (d) of Figure~\ref{figure:fit}. With the same method as the iron line, the detection significance of the cyclotron resonant absorption feature was also at greater than 99.99\% confidence level.  The best-fit spectral parameters for the ObsID 93032-30-01-03 are listed in Table~\ref{table:fit}, together with the MJD, the observational date, the exposure time, and the model flux in 3--20~keV at a distance of 3.9 kpc \citep{McBride06}. The errors were estimated at the 90\% confidence level ranges of the parameters.

\section{Discussion}\label{Disscussion}

We have analyzed the optical and X-ray observations of MXB~0656$-$072 during different X-ray states. A 101.2 day period has been inferred from the \emph{RXTE}/ASM and \emph{Swift}/BAT lightcurves thanks to a series of X-ray outbursts which occurred between MJD~54350 and MJD~54850. Such a 101.2 day period was very likely the orbital period of MXB~0656$-$072. The position of this 101.2 day period and its 160.4~s spin period in the Corbet diagram support the Be/X-ray binary nature of the system. MXB~0656$-$072 has similar orbital parameters as the Be/X-ray binary A 0535+26, which has a 111-day orbital period and a 103-s spin period, and they almost occupy the same position in the Corbet diagram (see Figure~\ref{figure:corbet}). The folded \emph{RXTE}/ASM and \emph{Swift}/BAT light curves of MXB 0656$-$072, with an orbital period of $P_{orb}=101.2$~days, show a complex outburst profile (see Figure~\ref{figure:foldlc}). The duration of the outburst covers nearly half of one orbital phase ($\sim$ 50 days).  Using the third Kepler law, we can estimate the semi-major axis of the binary system with $a^3=P_{\mathrm{orb}}^2\times G(M_*+M_{\mathrm{X}})/4\pi^2$, where $M_*$ and $M_{\mathrm{X}}$ are the masses of the optical companion and neutron star, respectively. Since MXB~0656$-$072 (O9.7\,Ve) and A0535$+$26 (O9.7\,IIIe) have similar optical counterparts, we adopt $M_*=20\ M_{\sun}$ and $R_*=15\ R_{\sun}$ \citep{Grundstrom07}. The mass of the compact object is fixed to $M_{\mathrm{X}}=1.4\ M_{\sun}$, typical of a neutron star. These values yield $a\sim 253\ R_{\sun}=16.9\ R_*$. The H$\alpha$ emission line is generally found to arise in a region about 10$R_*$ from the central Be star \citep{Slettebak92}. In order to capture material and cause an X-ray outburst, the neutron star should pass within this region at the periastron point. Hence, the separation between the neutron star and the Oe star at the periastron point should be smaller than the H$\alpha$ emission region, given by $a(1-e)$$\sim$10$R_*$. Thus, we can estimate that the orbital eccentricity of MXB~0656$-$072 should be $e\sim0.4$, which is comparable to the eccentricity of A0535+26, i.e., $e$=0.47 \citep{Finger96}.

We have followed MXB~0656$-$072 spectroscopically since 2005 and also photometrically since 2007 at optical wavelengths at Xionglong Station, NAOC. The strong H$\alpha$  emission line showed a maximum EW of $\sim -25$~{\AA} during our 2007 observations, which were taken just before the X-ray outburst in 2007 November. A strong HeI~$\lambda$6678 line was also observed during the 2007 observations. The photometric observations showed that the $UBV$ magnitudes during our 2007 observations were about 0.2 mag larger with respect to those in our following observations (see Table~\ref{table:diffmag} and Figure~\ref{figure:uvb}). Spectroscopy of 2007 November 16 (during the rising phase of the first X-ray outburst) indicated that the H$\alpha$ still had a very strong emission during the X-ray outburst. It is generally believed that the H$\alpha$ emission line in Be stars is formed in the entire circumstellar disk \citep{Slettebak92}, while only the innermost part of the disk contributes significantly to the continuum flux. Due to the higher ionization potential energy, the formation region of the HeI~$\lambda$6678 line should be smaller than the nearby continuum region \citep{Stee98}. The consistency among the variability of the brightness in the $UBV$ bands in MXB~0656$-$072 suggests that the changes of the optical continuum emission might be caused by the physical changes in the circumstellar disk around the Oe star \citep{Janot-Pacheco87}. The strong H$\alpha$ emission corresponds to a more extended circumstellar disk, while the decrease of the optical brightness indicates the dilution of the inner disk. Moreover, the increase of the HeI~$\lambda$6678 emission in the 2007 observations should be connected with the formation of a denser inner disk.

To explain all the observational phenomena, a unified physical model should be assumed in MXB~0656-072. Observational results \citep{Rivinius01} and theoretical calculations \citep{Meilland06} suggest that after an outburst a low-density region seems to develop around the Be star. \citet{Rivinius01} and \citet{Meilland06} suggested that the outburst might be connected with the increased mass loss or mass ejection from the Be star. Some weeks to months after the outburst, the stellar radiation pressure gradually excavates the inner part of the disk and a low-density region could develop around the Be star and slowly grow outward \citep{Rivinius01}. With the vacuation of the inner disk, the optical continuum emission decreases and an increase in $UBV$ magnitudes will be observed. After the outburst, material is transferred into the disk and a more extended circumstellar disk should be formed, which produces the stronger H$\alpha$ emission from the system. Therefore, we suggest that a mass ejection event had taken place in MXB~0656-072 before our 2007 observations and a low-density region was developing during that period. The HeI~$\lambda$6678 became stronger when the optical continuum emission was decaying, which indicated that a larger quantity of material close to the stellar surface should be present when a low-density region was developing in the inner part of the disk. The new disk material might be ejected from the star by the subsequent mass ejection or reaccreted into the inner region after the supply of material from the star to the disk has been turned off \citep{Clark01}.

\begin{center}
\begin{figure}
\centering
\includegraphics[width=12cm]{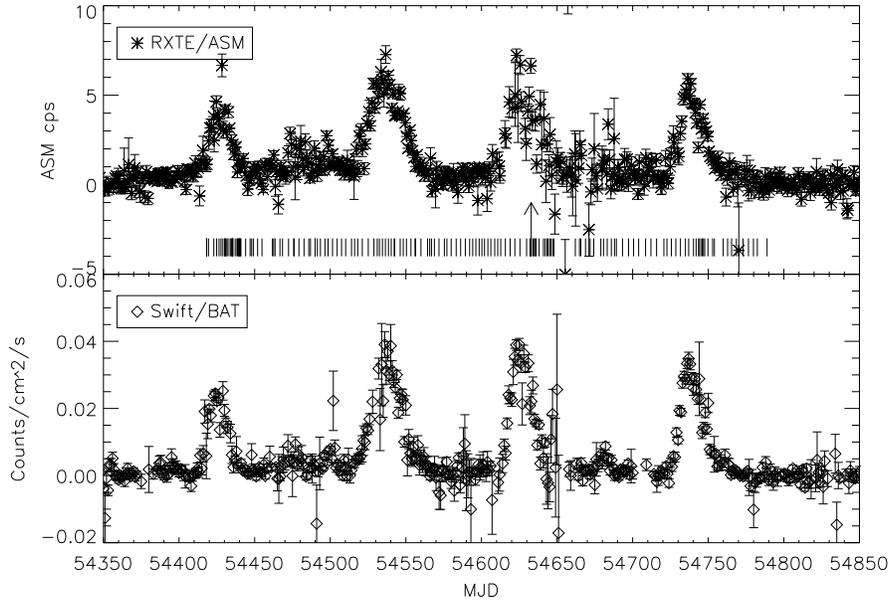}
\caption{\emph{RXTE}/ASM and \emph{Swift}/BAT light curves displaying several outbursts. The vertical bars in the top panel correspond to the dates of the RXTE-pointed observations. The arrow indicates the date of the ObsID~93032-30-01-03.}
\label{figure:asmbatlc}
\end{figure}
\end{center}

With the expansion of the circumstellar disk after the outburst, the outer part of the disk interacts with the neutron star around its periastron passage. A part of the material in the disk is accreted onto the neutron star and an X-ray outburst is triggered. This could explain the first X-ray outburst in 2007 which occurred in a decline phase of the optical brightness. A considerable fraction of the angular momentum would be transferred onto the neutron star in the system, which may account for the simultaneous spin-up detected by Fermi/Gamma-ray Burst Monitor\footnote{http://gammaray.msfc.nasa.gov/gbm/science/occultation/} during the last X-ray outburst in Figure~\ref{figure:asmbatlc}. A similar phenomenon was also observed in other Be/X-ray binaries, such as 4U~1145-619 \citep{Stevens97}, A0535+26 \citep{Clark99,Yan11} and 4U~0115+63 \citep{Reig07}. All these X-ray outbursts seem to take place in the decline phase of the optical emission.

During the first X-ray outburst in 2007, the H$\alpha$ emission kept nearly unchanged. The motion of the neutron star could not influence the size of the H$\alpha$ emission region instantaneously. With the extension of the ejected material, the density in the outer region of the disk would increase and a much stronger peak flux was observed in the second and third X-ray outbursts (see Figure~\ref{figure:asmbatlc}). The last X-ray outburst in 2008 occurred between MJD~54720 and MJD~54760. Our 2008 optical observations taken during the last X-ray outburst show that the H$\alpha$ emission level and the $UBV$ brightness returned to their levels of 2006. There were no more changes in H$\alpha$ emission and the $UBV$ brightness between our 2008 and 2009 observations. However, no X-ray outburst has been observed since the last one in 2008,indicating that the material ejected in 2007 has been dispersed into the outer region of the system. The circumstellar disk in the equatorial region of Oe star became denser and larger at the early stage of the mass ejection in the end of 2007 and a stronger H$\alpha$ emission was observed during our 2007 observations. Due to the gravitational effect of the neutron star, the size of the circumstellar disk could not increase continuously and it should have been truncated \citep{Okazaki01}. The size of the truncated disk after the last X-ray outburst in 2008 should be smaller than the distance to the first Lagrangian (L1) point at periastron and the H$\alpha$ emission was in a stable level after our 2008 observations.

Another large, extended Type II X-ray outburst was observed with \emph{RXTE} in MXB~0656$-$072 from 2003 October to 2004 January \citep{McBride06}. The duration of that Type II X-ray outburst is longer than the X-ray outbursts observed between 2007 November and 2008 November. It lasted about 100 days between MJD~52924 and MJD 53026. There were no systematic optical observations during 2003 X-ray outburst and we could not know what happened in MXB~0656$-$072. According to its long X-ray outburst duration (nearly the whole orbital period), we speculate that a much stronger mass ejection might have taken place before its 2003 outburst. Due to the spectacular changes of the stellar wind environment during the Type II X-ray outburst, the disk size would increase very rapidly and the outer disk material could interact with the neutron star before or after it approaches to the periastron point. Therefore, the Type II X-ray outburst could occur at any orbital phase and last a longer time than the Type I X-ray outburst.

The PCA$+$HEXTE spectra of MXB~0656$-$072 during the 2007-2008 X-ray outbursts could be well fitted by a cut-off power law model with a low-energy absorption, together with an iron fluorescent line at $\sim$6.4~keV and a cyclotron feature at $\sim$30.2~keV. Unlike the model suggested by \citet{McBride06} for the X-ray spectrum during the 2003 X-ray outburst, a blackbody component is not necessary in our model for the X-ray spectrum during the 2007-2008 X-ray outbursts. The detection significance for the iron line around 6.4~keV and the cyclotron resonance scattering feature around 30~keV is at a larger than 99.99\% confidence level. The X-ray luminosity in the 3-22 keV could be calculated as $6.6\times10^{36}~erg~s^{-1}$, assuming a distance of 3.9 kpc \citep{McBride06}. Once the energy of the cyclotron resonance line was known, the strength of the magnetic field in the surface of the neutron star could be estimated as
 \begin{equation}
\label{euqation3}
B=\frac{E_{cycl}}{11.6keV}\times(1+z)\times10^{12}G,
\end{equation}
where $E_cycl$ is the energy of the fundamental cyclotron resonant scattering line and $z$ is the gravitational redshift, respectively \citep{Coburn02}. Assuming $E_{cycl}$ = 30.2$^{+1.9}_{-2.9}$ in Table~\ref{table:fit} is the fundamental cyclotron line and $z$ = 0.3 for a typical neutron star with mass of 1.4~$M_{\sun}$ and radius of 10~km \citep{McBride06}, the magnetic field could be calculated by Equation~(\ref{euqation3}) to be $3.4^{+0.2}_{-0.3}\times10^{12}$G. The same variability of the soft and hard colors during the X-ray outbursts suggest no overall changes in the spectral shape of MXB 0656-072 in the 2.3-21.0 keV band. The X-ray flux during the outbursts should only be connected with the changes of the mass-accretion rate onto the neutron star around the periastron point.

\section{Conclusions}\label{SecSum}

We presented and analyzed the optical and X-ray observations of MXB~0656$-$072 during different X-ray outbursts.
A 101.2 day orbital period was found from a series of X-ray outbursts of the system between MJD~54350 and MJD~54850. The position of this orbital period and the 160.4~s spin period in the Corbet diagram confirms the Be/X-ray binary nature of MXB~0656$-$072. The anti-correlation between the H$\alpha$ emission and the $UBV$ brightness during our 2007 observations indicates that a mass ejection event might have taken place before the first X-ray outburst in 2007 November. A low-density region was 
developed around the Oe star in MXB~0656$-$072 after the mass ejection event, which could explain the decrease of the optical brightness during the 2007 observations. Material was transferred into the circumstellar disk during the mass ejection event and a more extended disk formed after the mass outburst, which should be the reason for the strong H$\alpha$ emission before the 2007 X-ray outburst. With the outward motion of the disk, the neutron star interacted with the outer part of the disk around the periastron passage and a series of transient X-ray outbursts could be triggered. According to the long duration of the 2003 Type II X-ray outburst, we speculate that a much stronger mass ejection event should have taken place before the 2003 outburst. The ejected material could fill all the orbital regions of the neutron star.

The PCA-HEXTE spectra during the 2007-2008 X-ray outbursts could be well fitted by a cut-off power law with low energy absorption. A Gaussian iron line around 6.4~keV and a cyclotron resonance scattering line around 30~keV are also considered in the model. The same variability of the two X-ray colors in the 2.3-21~keV band indicated no overall changes in the spectral shape during the X-ray outbursts, which might only be connected with the changes of the mass-accretion rate onto the neutron star.

\acknowledgements
We thank the anonymous referee for insightful suggestions. This work was supported by the Centre National d'Etudes Spatiales (CNES), the National Basic Research Program of China - 973 Program 2009CB824800, the Natural Science Foundation of China under Grants 10873036 and 11003045, the National High Technology research and Development Program of China - 863 project 2008AA12Z304, the Young Researcher Grant of Purple Mountain Observatory, CAS, and the Open Project program of Key Laboratory of Optical Astronomy, NAOC, CAS. It is based on observations obtained through MINE: the Multi-wavelength INTEGRAL NEtwork. J.Z.Y. is grateful to Liming Song and Yupeng Chen at IHEP, CAS, to Felix Mirabel, Stephane Corbel, Jerome Rodriguez, Farid Rahoui, and Tao Chen at CEA Saclay, for their kind help while my visiting Beijing and Paris, respectively, to Michael Nowak, Isabel Caballero, and Manfred Hanke for their valuable discussions, and to Yizhong Fan for his improvement of our manuscript. J.A.Z.H. acknowledges the Swiss National Science Foundation for financial support.

\begin{deluxetable}{ccccc}
\tablecolumns{9}
\tablecaption{Summary of the spectroscopic observations of MXB~0656$-$072.}
\tablewidth{0pc}
\tablehead{\colhead{Date (YYYYMMDD)}&\colhead{MJD} & \colhead{Exposure (s)} &\colhead{EW (--\AA)} & \colhead{Error (\AA)}}
 \startdata
20051023 & 53666.88 & 1000.0 & 14.44 & 0.13\\
20051023 & 53666.89 & 1000.0 & 14.20 & 0.38\\
20051024 & 53667.89 & 1000.0 & 14.05 & 0.16\\
20060928 & 54006.85 & 1200.0 & 16.81 & 0.13\\
20060929 & 54007.85 & 1200.0 & 16.02 & 0.31\\
20061001 & 54009.88 & 1200.0 & 17.03 & 0.21 \\
20061002 & 54010.88 & 900.0   & 17.14 & 0.32\\
20071028 & 54401.91 & 120.0   & 24.34 & 0.45\\
20071030 & 54403.88 & 1800.0 & 25.06 & 0.30 \\
20071031 & 54404.85 & 1800.0 & 25.25 & 0.16\\
20071101 & 54405.87 & 1800.0 & 25.41 & 0.34 \\
20071116 & 54420.77 & 1800.0 & 23.75 & 0.23 \\
20071116 & 54420.84 & 2400.0 & 24.16 & 0.22\\
20080930 & 54739.85 &1800.0  & 16.88 & 0.36\\
20081001 & 54740.84 & 1800.0 & 16.38 & 0.31\\
20081005 & 54744.86 & 1800.0 & 16.19 & 0.09\\
20081006 & 54745.85 & 1800.0 & 16.35 & 0.18\\
20081009 & 54748.86 & 1800.0 & 16.38 & 0.09\\
20091023 & 55127.90 & 1200.0 & 16.92 & 0.34\\
20091024 & 55128.87 & 1500.0 & 16.96 & 0.14\\
20091025 & 55129.86 & 1800.0 & 15.97 & 0.16\\
20091028 & 55132.90 & 1800.0 & 17.91 & 0.12\\
20091029 & 55133.86 & 2100.0 & 17.73 & 0.11\\
\enddata
\label{table_halpha}
\end{deluxetable}

\begin{deluxetable}{lccc}
\tablecolumns{3}
\tablecaption{Log of the $UBV$ optical photometric observations of MXB~0656$-$072.}
\tablewidth{0pc}
\tablehead{\colhead{Date} & \colhead{Exposures [$U/B/V$]} & \colhead{Seeing (arcsec)}}
\startdata
2007 Oct 30 & 480s+360s/240s+120s/2$\times$120s & 2.0 \\
2007 Oct 31 & 360s/120s/60s & 2.0\\
2007 Nov 1 & 2$\times$360s/2$\times$120s/2$\times$60s & 3.5\\
2008 Oct 5 &  3$\times$480s/3$\times$120s/3$\times$60s & 3.5\\
2008 Oct 6 &  3$\times$480s/3$\times$120s/3$\times$60s & 3.5\\
2008 Oct 8 & 4$\times$480s/4$\times$120s/2$\times$60s & 2.0\\
2009 Oct 27 & 4$\times$540s/4$\times$120s/4$\times$60s & 4.7\\
2009 Oct 28 & 2$\times$540s/2$\times$120s/2$\times$60s & 2.3\\
\enddata
\label{table:logphotometry}
\end{deluxetable}

\begin{deluxetable}{lcccccc}
\tiny
\tablecolumns{7}
\tablecaption{$UBV$ Differential Magnitudes of MXB~0656$-$072 and the Check Star \#4 Using the Main Reference Star \#3.}
\tablewidth{0pc}
\tablehead{\colhead{MJD} & \colhead{$\Delta$U} & \colhead{$\Delta$U (\#4)} &  \colhead{$\Delta$B} &  \colhead{$\Delta$B (\#4)} & \colhead{$\Delta$V} & \colhead{$\Delta$V (\#4)} }
\startdata
54403.8&0.727$\pm$0.006&0.942$\pm$0.007&0.887$\pm$0.002&0.911$\pm$0.002&0.449$\pm$0.001&0.960$\pm$0.002\\
54403.8&0.734$\pm$0.006&0.944$\pm$0.007&0.893$\pm$0.002&0.914$\pm$0.002&0.455$\pm$0.001&0.959$\pm$0.002\\
54404.9&0.728$\pm$0.006&0.938$\pm$0.007&0.887$\pm$0.003&0.906$\pm$0.003&0.449$\pm$0.003&0.958$\pm$0.004\\
54405.8&0.735$\pm$0.005&0.942$\pm$0.006&0.886$\pm$0.002&0.914$\pm$0.002&0.444$\pm$0.002&0.958$\pm$0.002\\
54405.8&0.727$\pm$0.005&0.934$\pm$0.006&0.882$\pm$0.002&0.911$\pm$0.002&0.442$\pm$0.002&0.957$\pm$0.002\\
54744.9&0.520$\pm$0.003&0.983$\pm$0.004&0.710$\pm$0.001&0.911$\pm$0.002&0.213$\pm$0.001&0.959$\pm$0.002\\
54744.9&0.530$\pm$0.003&0.976$\pm$0.004&0.709$\pm$0.001&0.912$\pm$0.002&0.214$\pm$0.001&0.959$\pm$0.002\\
54744.9&0.530$\pm$0.003&0.980$\pm$0.004&0.718$\pm$0.002&0.916$\pm$0.002&0.219$\pm$0.001&0.963$\pm$0.002\\
54745.8&0.545$\pm$0.004&0.976$\pm$0.004&0.723$\pm$0.002&0.913$\pm$0.002&0.230$\pm$0.001&0.955$\pm$0.002\\
54745.8&0.540$\pm$0.004&0.968$\pm$0.004&0.731$\pm$0.002&0.915$\pm$0.002&0.231$\pm$0.001&0.954$\pm$0.002\\
54745.8&0.546$\pm$0.004&0.967$\pm$0.004&0.718$\pm$0.002&0.914$\pm$0.002&0.233$\pm$0.001&0.960$\pm$0.002\\
54747.8&0.532$\pm$0.004&0.976$\pm$0.004&0.715$\pm$0.002&0.919$\pm$0.002&0.223$\pm$0.001&0.948$\pm$0.002\\
54747.8&0.540$\pm$0.004&0.976$\pm$0.004&0.728$\pm$0.003&0.917$\pm$0.004&0.235$\pm$0.001&0.958$\pm$0.002\\
54747.8&0.538$\pm$0.003&0.973$\pm$0.004&0.716$\pm$0.002&0.910$\pm$0.002&0.232$\pm$0.001&0.958$\pm$0.002\\
54747.8&0.532$\pm$0.003&0.966$\pm$0.004&0.722$\pm$0.002&0.919$\pm$0.002&0.228$\pm$0.001&0.960$\pm$0.002\\
55131.8&0.568$\pm$0.035&0.900$\pm$0.042&0.784$\pm$0.008&0.929$\pm$0.008&0.291$\pm$0.009&0.986$\pm$0.013\\
55131.8&0.579$\pm$0.017&1.039$\pm$0.023&0.777$\pm$0.012&0.921$\pm$0.013&0.315$\pm$0.006&0.980$\pm$0.008\\
55131.9&0.598$\pm$0.011&0.964$\pm$0.013&0.754$\pm$0.007&0.914$\pm$0.007&0.281$\pm$0.004&0.957$\pm$0.007\\
55131.9&0.592$\pm$0.010&0.981$\pm$0.013&0.763$\pm$0.008&0.911$\pm$0.009&0.298$\pm$0.005&0.968$\pm$0.007\\
55132.9&0.581$\pm$0.005&0.985$\pm$0.006&0.767$\pm$0.004&0.919$\pm$0.004&0.291$\pm$0.003&0.967$\pm$0.004\\
55132.9&0.578$\pm$0.005&0.984$\pm$0.006&0.773$\pm$0.004&0.924$\pm$0.004&0.291$\pm$0.003&0.965$\pm$0.004\\
\enddata
\label{table:diffmag}
\end{deluxetable}

\begin{table}
\small	
\begin{center}
\caption{Best-fit Parameters for PCA-HEXTE Spectra of MXB~0656$-$072.}
\begin{tabular}{lc}
\tableline
Parameter & Value\tablenotemark{a}\\
\tableline
ObsID                 & 93032-30-01-03   \\
MJD                   & 54632.88        \\
Date                  & 2008 Jun 15   \\
Exposure (s)           & 4912       \\
$N_H$($10^{22}~cm^{-2}$) & 1.50$^{+0.53}_{-0.50}$ \\
$E_{Fe}$(keV)         & 6.44$^{+0.05}_{-0.03}$     \\
$Fe_{Norm}$\tablenotemark{b}         & 4.3$^{+0.4}_{-0.4}$    \\
$E_{GABS}$ (keV)              & 10.45$^{+0.22}_{-0.25}$      \\
$\sigma_{GABS}$ (keV)              & 2.83$^{+0.64}_{-0.53}$     \\
$\tau_{GABS}$              & 1.03$^{+0.77}_{-0.41}$    \\
$\Gamma$              & 0.37$^{+0.18}_{-0.19}$     \\
$E_{fold}$ (keV)         & 9.45$^{+2.02}_{-1.53}$    \\
$\alpha$\tablenotemark{c}            &0.10$^{+0.02}_{-0.02}$	  \\
$E_{cycl}$ (keV)          &30.2$^{+1.9}_{-2.9}$	  \\
$D_{cycl}$            &0.5$^{+0.3}_{-0.2}$	    \\
Factor\tablenotemark{d}            &0.59$^{+0.01}_{-0.01}$	        \\
$\chi^2_{red}$(dof)   & 1.01(46)           \\
Luminosity (3--22~keV)\tablenotemark{e}  & 6.6                            \\
\tableline
\tableline
\end{tabular}
\tablenotetext{a}{All errors represent 90\% confidence intervals. }
\tablenotetext{b}{$10^{-3}~phontons~cm^{-2}~s^{-1}$. }
\tablenotetext{c}{$phontons~keV^{-1}~cm^{-2}~s^{-1}$ at 1~keV. }
\tablenotetext{d}{The normalization constant for the HEXTE data.}
\tablenotetext{e}{In unit of $10^{36}~erg~s^{-1}$ for a distance of 3.9 kpc \citep{McBride06}.}
\label{table:fit}
\end{center}

\end{table}

\begin{thebibliography}{}


\bibitem[Barthelmy(2000)]{Barthelmy00}
Barthelmy, S. D., 2000, Proc. SPIE, 4140, 50


\bibitem[Clark et al.(1975)]{Clark75}
Clark, G. W., Schmidt, G. D. and Angel, J. R. P., 1975, IAU Circ., 2843

\bibitem[Clark et al.(1999)]{Clark99}
Clark, J.S., Lyuty, V.M., Zaitseva, G.V., et al., 1999, MNRAS, 302, 167

\bibitem[Clark et al.(2001)]{Clark01}
Clark, J.S., Tarasov, A.E., Okazaki, A.T., et al., 2001, \aap, 380, 615

\bibitem[Coburn et al.(2002)]{Coburn02}
Coburn, W., Heindl, W.A., et al., 2002, \apj, 580, 394


\bibitem[Corbet(1986)]{Corbet86}
Corbet, R. H. D., 1986, MNRAS, 220, 1047


\bibitem[Finger et al.(1996)]{Finger96}
Finger, M. H., Wilson, R. B., Harmon, B. A., 1996, \apj, 459, 288

\bibitem[Grundstrom et al.(2007)]{Grundstrom07}
Grundstrom, E. D., Boyajian, T. S., Finch, C., Gies, D. R., et al., 2007, \apj, 660, 1398

\bibitem[Houck \& Demicola(2000)]{Houck2000}
Houck, J.C. and Demicola, L.A., Astronomical Data Analysis Software and Systems IX, ASP Conference Proceedings, Vol. 216, edited by Nadine Manset, Christian Veillet, and Dennis Crabtree. Astronomical Society of the Pacific, 591

\bibitem[Janot-Pacheco et al.(1987)]{Janot-Pacheco87}

Janot-Pacheco, E., Motch, C., \& Mouchet, M., 1987, \aap, 177, 91

\bibitem[Kaluzienski et al.(1976)]{Kaluzienski76}
Kaluzienski, L. J., Holt, S. S., Boldt, E. A., Serlemitsos, P. J. and Bortle, J., 1976, IAU  Circ., 2935

\bibitem[Kennea et al.(2007)]{Kennea07}
Kennea, J. A., Romano, P., Pottschmidt, K., Wilms, J., et al., 2007, ATel, 1293

\bibitem[Kreykenbohm et al.(2007)]{Kreykenbohm07}
Kreykenbohm, I., Shaw, S. E., Bianchin, V.,  Diehl, R., et al., 2007, ATel, 1281


\bibitem[Levine et al.(1996)]{Levine96}
Levine, A. M., Bradt, H., Cui, W., Jernigan, J. G., et al., 1996, ApJ, 469, L33

\bibitem[Liu et al.(2001)]{Liu01}
Liu, Q. Z., van Paradijs, J., and van den Heuvel, E. P. J., 2001, \aap, 368, 1021

\bibitem[Liu et al.(2006)]{Liu06}
Liu, Q. Z., van Paradijs, J., and van den Heuvel, E. P. J., 2006, \aap, 455, 1165

\bibitem[Makishima et al.(1990)]{Makishima90}
Makishima, K., Ohashi, T., Kawai, N., et al., 1990, PASJ, 42, 295

\bibitem[McBride et al.(2006)]{McBride06}
McBride, V. A., Wilms, J., Coe, M. J., et al., 2006, \aap, 451, 267


\bibitem[Meilland et al.(2006)]{Meilland06}
Meilland, A., Stee, Ph., Zorec, J., and Kanaan, S., 2006, \aap, 455, 953

\bibitem[Mihara et al.(1990)]{Mihara90}
Mihara, T., Makishima, K., Ohashi, T., et al., 1990, Nature, 346, 250


\bibitem[Morgan et al.(2003)]{Morgan03}
Morgan, E., Remillard, R., and Swank, J., 2003, ATel, 199
\bibitem[Okazaki \& Negueruela(2001)]{Okazaki01}
Okazaki, A. T. and Negueruela, I., 2001, \aap, 377, 161

\bibitem[Pakull et al.(2003)]{Pakull03}
Pakull, M., Motch, C., and Negueruela, I., 2003, ATel, 202

\bibitem[Porter \& Rivinius(2003) ]{Porter03}
Porter, J. M. and Rivinius, T., 2003, PASP, 115, 1153

\bibitem[Pottschmidt et al.(2007)]{Pottschmidt07}
Pottschmidt, K., McBride, V. A., Suchy, S., Kreykenbohm, I., et al., 2007, ATel, 1283

\bibitem[Protassov et al.(2002)]{Protassov02}
Protassov, R., van Dyk, D.A., Connors, A., et al., 2002, \apj, 571, 545

\bibitem[Raguzova \& Popov(2005)]{Raguzova05}
Raguzova, N. V. and Popov, S. B., 2005, A\&AT, 24, 151

\bibitem[Reig(2011)]{Reig11}
Reig, P., 2011, Astrophysics and Space Science, 332, 1

\bibitem[Reig et al.(2007)]{Reig07}
Reig, P., Larionov, V., Negueruela, I., et al., 2007, \aap, 462, 1081

\bibitem[Remillard \& Marshall(2003)]{Remillard03}
Remillard, R. and Marshall, F., 2003, ATel, 197

\bibitem[Rivinius et al.(2001)]{Rivinius01}
Rivinius, Th., Baade, D., Stefl, S., and Maintz, M., 2001, \aap, 379, 257



\bibitem[Scargle(1982)]{Scargle82}
Scargle, J. D., 1982, \apj, 263, 835

\bibitem[Slettebak et al.(1992)]{Slettebak92}
Slettebak, A., Collins, G.W., II, Truax, R., 1992, ApJS, 81, 335


\bibitem[Stee et al.(1998)]{Stee98}
Stee, Ph., Vakili, D., Bonneau, D., \& Mourard, D., 1998, \aap, 332, 268


\bibitem[Stevens et al.(1997)]{Stevens97}
Stevens, J.B., Reig, P., Coe, M.J., et al., 1997, MNRAS, 288, 988

\bibitem[White et al.(1983)]{White83}
White, N.E., Swank, J.H., \& Holt, S.S., 1983, \apj, 270, 711

\bibitem[Yan, Liu \& Li(2007)]{Yan07a}
Yan, J.Z., Liu, Q.Z., and Li, H., 2007, ATel, 1303

\bibitem[Yan et al.(2012)]{Yan11}
Yan, J.Z., Li, H., and Liu, Q.Z., 2012, ApJ, 744, 37

\bibitem[Zhang et al.(2004)]{Zhang04}
Zhang, F., Li, X.-D. and Wang, Z.-R., 2004, \apj, 603, 663


\end{thebibliography}
\end{document}